\documentclass[usenatbib]{mnras}

\usepackage{newtxtext,newtxmath}
\usepackage{xspace}
\usepackage{bm}

\usepackage{float}
\usepackage[T1]{fontenc}
\usepackage{amsmath}
\usepackage{xspace}

\DeclareRobustCommand{\VAN}[3]{#2}
\let\VANthebibliography\thebibliography
\def\thebibliography{\DeclareRobustCommand{\VAN}[3]{##3}\VANthebibliography}

\newcommand{\LODI}{\texttt{LODI}\xspace}
\newcommand{\mpc}{$\,h^{-1}$Mpc\xspace}
\newcommand{\FOLD}{\texttt{Cosmo-FOLD}\xspace}
\newcommand{\FOLDpos}{\texttt{Cosmo-FOLD+pos}\xspace}
\newcommand{\Star}{$\boldsymbol{\rho}_\text{star}$\xspace}
\newcommand{\dm}{$\boldsymbol{\rho}_\text{DM}$\xspace}
\newcommand{\Tg}{$\textbf{T}_g$\xspace}
\newcommand{\field}[3]{#1, #2 Mpc $h^{-1}$, #3$^3$} 
\newcommand{\camels}{\texttt{CAMELS}\xspace}
\newcommand{\tng}{\texttt{TNG300-2}\xspace}
\newcommand{\cfield}{\xi} 
\newcommand{\ofield}{\eta} 
\newcommand{\NNpars}{\boldsymbol{\phi}} 
\newcommand{\tem}{\boldsymbol{\tau}^t}
\newcommand{\unitymat}{\mathbb{I}}

\usepackage{graphicx}	
\usepackage{amsmath}	
\usepackage{algorithm}
\usepackage{algpseudocode}
\algnewcommand{\HashComment}[1]{\hfill \texttt{\# #1}}

\title[\FOLD]{\FOLD: Fast generation and upscaling of field-level cosmological maps with overlap latent diffusion}

\author[Mishra, Trotta \& Viel]{
Satvik Mishra,$^{1,2}$\thanks{E-mail: samishr@sissa.it}
Roberto Trotta,$^{1,3,4,5}$
Matteo Viel$^{1,2,3,4,6,7}$
\\
$^{1}$Theoretical and Scientific Data Science, SISSA, Via Bonomea 265, 34136 Trieste, Italy\\
$^{2}$Astroparticle and Gravitational Physics Group, SISSA, Via Bonomea 265, 34136 Trieste, Italy\\
$^{3}$INFN -- National Institute for Nuclear Physics, Via Valerio 2, 34127 Trieste, Italy\\
$^{4}$ICSC - Centro Nazionale di Ricerca in High Performance Computing, Big Data e Quantum Computing, Via Magnanelli 2, Bologna, Italy\\
$^{5}$Astrophysics Group, Physics Department, Blackett Lab, Imperial College London, Prince Consort Road, London SW7 2AZ, UK\\
$^{6}$INAF -- Osservatorio Astronomico di Trieste, Via G. B. Tiepolo 11, I-34143 Trieste, Italy\\
$^{7}$IFPU -- Institute for Fundamental Physics of the Universe, Via Beirut 2, I-34151 Trieste, Italy\\
}

\date{Accepted XXX. Received YYY; in original form ZZZ}

\pubyear{\the\year{}}

\begin{document}
\label{firstpage}
\pagerange{\pageref{firstpage}--\pageref{lastpage}}
\maketitle

\begin{abstract}
We demonstrate the capabilities of probabilistic diffusion models to reduce dramatically the computational cost of expensive hydrodynamical simulations to study the relationship between observable baryonic cosmological probes and dark matter at field level and well into the non-linear regime. We introduce a novel technique, \FOLD (Cosmological 
Fields via Overlap Latent Diffusion) to rapidly generate accurate and arbitrarily large cosmological and astrophysical 3-dimensional fields, conditioned on a given input field. We are able to generate \tng dark matter density and gas temperature fields from a model trained only on $\approx1\%$ of the volume (a process we refer to as `upscaling'), reproducing both large scale coherent dark matter filaments and power spectra to within $10\%$ for wavenumbers  $k\leq5$\mpc. These results are obtained within a small fraction of the original simulation cost and produced on a single GPU. Beyond one and two points statistics, the bispectrum is also faithfully reproduced through the inclusion of positional encodings. Finally, we demonstrate \FOLD's generalisation capabilities by upscaling a \camels volume of $(25\text{\mpc})^3$ to a full \tng volume of $(205 \text{\mpc})^3$ with no fine-tuning. \FOLD opens the door to full field-level simulation-based inference on cosmological scale.

\end{abstract}




\section{Introduction}

Cosmological simulations are essential tools for understanding the formation and evolution of structure in the Universe. They allow us to explore the complex interplay of gravitational, hydrodynamic, and astrophysical processes shaping the distribution of matter from large down to highly non-linear galactic and sub-galactic scales. However, high-fidelity hydrodynamical simulations are computationally very expensive, limiting their applicability when the parameter space is large  in inference studies. For example, the IllustrisTNG simulations~\citep{nelson2021illustristngsimulationspublicdata,Pillepich_2017} required approximately 35 million CPU-hours for the largest \tng run and about 18 million CPU-hours for TNG100, while EAGLE~\citep{EAGLE} consumed roughly 4.5 million CPU-hours.

Such computational demands become particularly prohibitive for cosmological or astrophysical field-level inference and simulation-based inference (SBI) frameworks, which require large ensembles of conditional forward simulations to explore the parameter spaces. Recent works have shown promising applications, including the cosmological context~\citep{angulo_large-scale_2022,jeffrey_likelihood-free_2020,bairagi2025simulationsneedsimulationbasedinference,Zeghal_2025,karchev2023simsimssimulationbasedsupernovaia,Saxena_2024,tucci24}. However, their applicability remains limited by the requirement of generating sufficient high-fidelity simulations. This computational bottleneck underlines the need for fast, accurate surrogate models capable of reproducing cosmological fields at a fraction of the cost -- not limited to dark matter, but crucially including baryonic observables that can be directly constrained with data. Recent works have explored various approaches to modeling cosmological fields using generative adversarial networks \citep[e.g.,][]{andrianomena2022}, probabilistic methods \citep{ono_debiasing_2024}, and diffusion-based techniques \citep[e.g.,][]{rouhiainen_cosmology_2024,mishra2025largefastaccuratehi}. These methods aim to overcome the computational bottlenecks and enable fast sampling of cosmic structures, preserving both large-scale coherence and small-scale physical accuracy. A related goal, currently hampered by similar computational bottlenecks, is to complement traditional summary statistics (such as two- and three-point statistics) with more informative data-driven summaries that best capture non-Gaussian features of the evolved non-linear fields (e.g.~\cite{bairagi2025patchnethierarchicalapproachneural,Borg,lanzieri2025optimalneuralsummarisationfullfield,Bayer_2025}.

In recent years, machine learning techniques, and in particular generative models, have emerged as powerful emulators to reproduce cosmological and astrophysical fields~\citep{perraudin2019cosmologicalnbodysimulationschallenge,Li_2021,jamieson_field-level_2023,Rigo_2025,hassan_hiflow_2022,andrianomena_emulating_2022,sharma25,bernardini2025ember2,andrianomena2024cosmologicalmultifieldemulator}. Among them, diffusion models—a class of stochastic generative models that generate data by learning to reverse a gradual noising process by iteratively denoising corrupted data samples—have gained significant attention due to their ability to produce high-fidelity, diverse outputs~\citep{sohl-dickstein_deep_2015, ho_denoising_2020, kingma_variational_2023}. These models have been successfully applied in astrophysics and cosmology, for example to emulate the baryonic processes affecting galaxy properties \citep{Chadayammuri_2023} or more in general to augment  the resolution of  simulations by using  a smaller set of high-resolution ones \citep{kodi20}.

Despite these advances, a major challenge remains in scaling generative models to arbitrarily large, high-fidelity cosmological volumes, especially when constrained by GPU memory and computational cost. Moreover, the evolution of cosmic structure is inherently non-local, where long-range gravitational interactions couple distant regions and baryonic feedback processes, such as those driven by supernovae or active galactic nuclei, can redistribute matter over large scales. Accurately modeling these effects requires the network to learn global dependencies rather than simple local conditional relationships, which becomes particularly challenging when no similar low-resolution reference field is available to guide the generation.

To overcome these limitations, we introduce \FOLD (for ``Cosmological Field Overlap Latent Diffusion''), a novel technique building on the variational diffusion model framework \LODI presented in~\cite{mishra2025largefastaccuratehi}. \FOLD enables the rapid and accurate generation of large three-dimensional cosmological fields by conditioning the denoising process on the underlying large scale matter distribution, represented by either the stellar density or dark matter density fields. We demonstrate the capability of our approach by generating gas temperature fields conditioned on dark matter density, as well as dark matter density fields reconstructed from sparse stellar density field distribution of IllustrisTNG--TNG300 simulations. 
These two applications are particularly challenging and relevant since they are both intrinsically sensitive to small scale physical processes (star formation and galactic feedback), which also affect intermediate cosmological scales. 
Another reason for the choice of these two applications and their importance is how they are manifested in terms of a summary statistics like the power spectrum. It is well known that baryonic processes, as simulated in state-of-the-art hydrodynamical simulations, determine a suppression of power at $k\sim~1\, h/$Mpc  caused by feedback, and a steep increase of power at smaller scales $k\gg 1\,h/$Mpc produced by stars forming in the innermost parts of the (dark matter) haloes \citep{chisari19,schneider15,vandaalen20}. While the exact amount of suppression and increase of power varies among the different sets of simulations, the shape of the effects is a common prediction of most of the models.
Our choices for the temperature field and stellar fields is a natural choice to address these two different scales and verify the performance of our method in these regimes.

The model is trained on just $\approx1\%$ and validated on $\approx0.5\%$ of the simulation volume, before being applied to synthesize complete large-scale realization, with excellent performance across 1-, 2- and 3-points statistics. \FOLD naturally enforces periodic boundary conditions and surpasses previous overlapping patch-based approaches by eliminating edge artifacts and maintaining structural coherence even when conditioned on the highly sparse stellar input.

The paper is structured as follows: in Section~\ref{sec:methods}, we describe the datasets used and the methodology. Section~\ref{sec:results} presents results on both test datasets and larger simulation volumes. We conclude with a discussion  in Section~\ref{sec:conclusions}.


\section{Methodology}
\label{sec:methods}

In this section, we present the machine learning framework and the simulation dataset used in this work. The driving idea is to generate a 3D target field (which could be a DM-only, neutral hydrogen or stellar mass density field), conditional on an input field by using  a variational diffusion model~\cite{kingma_variational_2023,mishra2025largefastaccuratehi}. These `small' fields (typically, cubes of 25 $h^{-1}$Mpc by side) are then upscaled to a much larger volume (typicaly, hundreds of Mpc by side) by using the novel \FOLD algorithm, designed to ensure that both large-scale structure and fine-scale physical processes are coherently represented.



\vspace{\baselineskip}

\subsection{Probabilistic diffusion model}
\begin{figure*}
    \centering
    \includegraphics[width=1.0\linewidth]{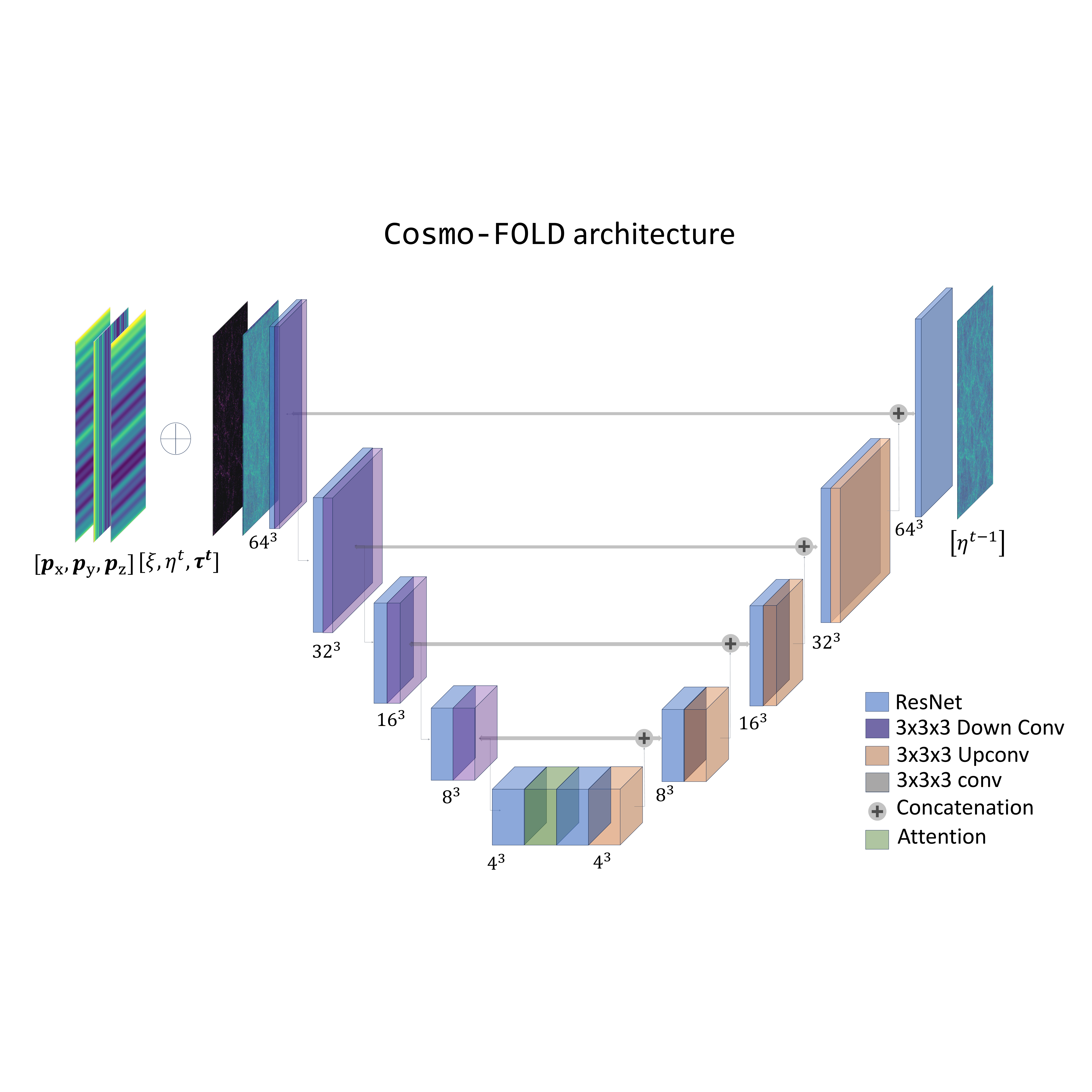}
    \caption{Architecture of the denoising network of \FOLD: a 3D U-net with residual network blocks and a self-attention bottleneck. The inputs to the network are the conditioning field $\cfield$, the noised target $\ofield^t$ at time step $t$, the sinusoidal time step embedding $\tem$ corresponding to the diffusion time step $t$, and the optional positional embedding $[\mathbf{p}_x, \mathbf{p}_y, \mathbf{p}_z]$ for each of the coordinate axes. The output $\ofield^{t-1}$ is the target field denoised by one extra step.}
    \label{fig:DenoiseUnet}
\end{figure*}

Diffusion models, proposed by~\cite{sohl-dickstein_deep_2015,ho_denoising_2020}, are a class of generative models that stochastically synthesize samples from an underlying data distribution by learning to reverse a gradual noising process applied to the data. The forward (so-called `noising') process progressively adds Gaussian noise to a data point $\textbf{x}_0$, yielding a series of latent and progressively more noisy representations  $\textbf{x}_t$, with $t \in \{1, 2, \dots, \text{T}\}$, where $T$ is the total number of noising steps. This process can be written as:
\begin{equation}\label{eq:Gaussian_noising}
    q\left( \textbf{x}_t|\textbf{x}_0 \right) = \mathcal{N}_{\textbf{x}_t} \left( \alpha_t\textbf{x}_0,\sigma^2_t \unitymat \right) \, ,
\end{equation}
 where $\mathcal{N}_x(\mathbf{\mu}, \mathbf{\Sigma})$ is a multivariate Gaussian in $\mathbf{x}$ with mean $\mathbf{\mu}$ and covariance matrix $\mathbf{\Sigma}$, and $\unitymat$ denotes the unity matrix. The quantities $\alpha_t$ and $\sigma_t$ are parametrized according to a noise schedule. In the reverse (or `denoising') process, the model learns to iteratively denoise the added noise in $T$ steps:
\begin{equation}\label{eq2}
    p_{\NNpars}\left( \textbf{x}_{t-1} \mid \textbf{x}_t \right) = \mathcal{N}_{\textbf{x}_{t-1}} \left( \bm{\mu}_{\NNpars}\left( \textbf{x}_t,t \right),\mathbf{\Sigma}_{\NNpars}\left( \textbf{x}_t,t \right)\right) \, ,
\end{equation}
where the covariance $\mathbf{\Sigma}_{\NNpars}$ and mean $\bm{\mu}_{\NNpars}$ are predicted by a denoising neural network with parameters $\NNpars$. The model is trained with objective to predict $\textbf{x}_{t-1}$ given $\textbf{x}_t$ via a loss function of the form \citep{kingma_variational_2023}: 
\begin{align}
\mathcal{L}_T(\mathbf{x}) &=  \mathbb{E}_{\epsilon, t} \left[ \left( f_{\NNpars}(\sigma_t,\alpha_t\right) \|\bm{\epsilon} - \hat{\bm{\epsilon}}_{\NNpars}(\mathbf{x}_t; t)\|_2^2 \right]
\end{align}
where  $\boldsymbol{\epsilon}\sim\mathcal{N}(\mathbf{0},\textbf{I})$ represents the noise injected into the data during the forward noising process defined in Eq.~\eqref{eq:Gaussian_noising}
and $\hat{\bm{\epsilon}_{\NNpars}}$ is the estimated noise. The expectation is taken over the diffusion time $t$, where $t\sim\mathcal{U}(1,T)$.

\vspace{\baselineskip}

We adopt the variational diffusion model of~\citet{kingma_variational_2023}, which was implemented in our previous work,~\cite{mishra2025largefastaccuratehi}, in an algorithm that we called \LODI. \LODI trains  a variational diffusion model to sample a target 3D field, $\ofield$, conditioned on a field $\cfield$, by sampling from the conditional distribution $p_{\NNpars}\left(\ofield \mid \cfield \right)$.  For this paper, we use the same  denoising neural network as \LODI: a U-Net ~\citep{ronneberger_u-net_2015} with an encoder-decoder convolutional neural network with residual connections as the convolutional blocks and an attention block at the bottleneck to capture both global and short-range correlations in the data. The encoder has 4 downsampling and upsampling residual blocks, and it is shown in Fig.~\ref{fig:DenoiseUnet}. The inputs to the network are the conditioning field $\cfield$, the noised target $\ofield^t$ at time step $t$, the sinusoidal time step embedding $\tem$, defined as:

\begin{equation}
\begin{aligned}
\tem
&= \bigl(\sin(\tilde{t}\,\boldsymbol{\omega}), \cos(\tilde{t}\,\boldsymbol{\omega})\bigr),
\qquad \tilde{t}=1000\,t,\ \ t\in[0,1], \\
\boldsymbol{\omega}
&= \bigl(\omega_k\bigr)_{k=0}^{\frac{d}{2}-1},
\qquad
\omega_k
= \exp\!\left(
-\log(\tau_{\max}) \, \frac{k}{\frac{d}{2}-1}
\right).
\end{aligned}
\end{equation}
where $\tau_{\max}=10^4$ and $d$ is the embedding dimension. For our work we use $d=64$.

This paper introduces improves to the \LODI approach in the form of optional positional encodings $\mathbf{p}$, defined in Eq.~\eqref{eq:pos_embedding}, that are concatenated channel-wise to the the inputs of the diffusion model, thereby adding 3 extra channels, one for each coordinate axis. The other key novelty is a different overlap method for the generation of volumes larger than the training box, which is explained in section~\ref{subsect:contgen}.

\subsection{Training datasets}
\label{sec:halo_to_HI}

For this work, we use two separate simulated datasets. 
The CAMELS–IllustrisTNG subset (henceforth \camels)~\citep{villaescusa-navarro_camels_2022} are a suite of cosmological simulations across which cosmological ($\Omega_m$ and $\sigma_8$) and astrophysical parameters (two supernova feedback parameters, $A_\text{SN1}$ and $A_\text{SN2}$ and two AGN feedback parameters, $A_\text{AGN1}$ and $A_\text{AGN2}$) are varied. More precisely:  $A_\text{SN1}$  represents a normalization factor for flux of the galactic wind feedback, relating the overall energy output to the star formation, while $A_\text{SN2}$  represents a normalization factor for the speed of the galactic winds; similarly $A_\text{AGN1}$ represents a normalization factor for the energy output of AGN feedback while $A_\text{AGN2}$ affects the specific energy of AGN feedback. 
We use the IllustrisTNG realizations with a comoving box of side 25\mpc, with $256^3$ gas and dark matter particles. We use the Cosmic Variance (CV) set, comprising of 27 realizations run with identical cosmologies and astrophysics. but different initial conditions. For our work, we choose simulations with $\Omega_m=0.3089,\sigma_8=0.8159$, and the four astrophysical parameters (collectively denoted by $\mathbf{A}$) all set to $1$; this choice represents the fiducial model both in terms of cosmological parameters and parameterization of AGN/SN feedback.

The second set of simulations is Illustris TNG300-2 (henceforth \tng)~\citep{nelson2021illustristngsimulationspublicdata}, a full hydrodynamical simulation 
run in a comoving volume of side 205\mpc, with $1250^3$ gas and dark matter particles, which we use to demonstrate what we call `upscaling', i.e. the ability of \FOLD to generate target fields for volumes much larger than those it has been trained on. These simulations share the same cosmological parameters of \camels and with a setup for the feedback which corresponds to the fiducial case of \camels ($\mathbf{A}=1$).

All 3D fields are taken at redshift $z=0$, where the density field is maximally non-linear and therefore the hardest to predict. From each simulation, we extract the stellar density \Star, cold dark matter density \dm and mass-weighted mean gas temperature \Tg (from now on called simply `gas temperature'). For \camels, the fields are voxelized to $256^3$ for \Tg and $128^3$ \dm, corresponding to spatial resolutions of $\approx 0.1$\mpc and $\approx 0.2$\mpc, respectively. For the \tng simulation, \Tg, \dm and stellar density field \Star, are discretized into $1024^3$ voxels, with a spatial resolution of $\approx 0.2$\mpc. These configurations are summarized in Table~\ref{tab:fields}.

\begin{table*}\label{summary_table}
\centering
\caption{\FOLD use cases presented in this paper, including details of box sizes and voxelization scheme for each. `Upscaling' refers to the $\approx$100-fold volume increase between the training and the generated simulation boxes. For each field, we indicate its physical content, its box side and the number of voxels. For the last case, training is on \camels and generation is for the \tng conditioning field. \label{tab:fields}}
\begin{tabular}{l l l l l l}
\hline
\textbf{Use case} & \textbf{Training field} &  \textbf{Conditioning field, $\cfield$} & \textbf{Output field, $\ofield$} & \textbf{Figures}\\
\hline
\multicolumn{5}{c}{\camels} \\ \hline
Gas temperature from DM &  \field{\Tg}{25}{256} & \field{\dm}{25}{256} & \field{\Tg}{25}{256} & Fig.~\ref{fig:cont_gen}\\

\hline
\multicolumn{5}{c}{\tng} \\ \hline
Gas temperature from DM with upscaling&  \field{\Tg}{51.25}{256} & \field{\dm}{205}{1024} & \field{\Tg}{205}{1024} & Figs.~\ref{fig:upscaling}, \ref{fig:TNG_DM_mstar_1}, \ref{fig:TNG_DM_mstar_2}\\
DM from stellar field with upscaling &  \field{\dm}{51.25}{256} & \field{\Star}{205}{1024} & \field{\dm}{205}{1024} & Figs.~\ref{fig:upscaling}, \ref{fig:TNG_DM_mstar_1}, \ref{fig:TNG_DM_mstar_2} \\
As above + positional encoding &  \field{\dm}{51.25}{256} & \field{\Star}{205}{1024} & \field{\dm}{205}{1024} & Figs.~\ref{fig:bispectra}, \ref{fig:bispectra_2} \\ 
\hline
\multicolumn{5}{c}{\camels $\rightarrow$ \tng transfer} \\ \hline
DM from stellar field with upscaling&  \field{\dm}{25}{128} & \field{\Star}{205}{1024} & \field{\dm}{205}{1024} & Figs.~\ref{fig:TNG_CAMELS_illus}, \ref{fig:TNG_CAMELS_PS}\\
\hline
\end{tabular}
\end{table*}

\subsubsection{Generating gas temperature maps from dark matter maps}\label{gasfromdm}

Our fist use case is to predict $\ofield=$\Tg from the $\cfield=$\dm field, in two different modalities: first, we show with \camels the performance of the model when the box size is the same between training and generation; second, we demonstrate with \tng the ability of our method to upscale to a 100 times larger volume the generative capability learnt from a small simulation box.

For \camels, we train on three CV simulations (all with the same cosmological and astrophysical parameters) with different random seeds, each voxelized at $256^3$. For \tng, we train on a $256^3$ voxels sub cube, corresponding to a box of side $51.25$\mpc, taken from the full \tng run. In both cases, we subdivide this $256^3$ training volume into 64 subvolumes, each encompassing $64^3$ voxels, which are used as individual samples for training. We note that subdividing into $64^3$ voxels for training and generating is carried out due to GPU memory limitations, as larger training volumes would not fit into the available GPU memory. Due to the upscaling capabilities of our method, this is however not a fundamental limitation, and actually makes the learning more efficient. Data augmentation is applied by random rotations and reflections to enforce rotational equivariance. This gives us $24$ different orientations per training sample. Since their density spans over five orders of magnitude, all fields are log-scaled prior to training. This results in $\sim24{,}600$ training samples for \camels and $\sim8{,}200$ for \tng.  

\subsubsection{Generating dark matter maps from stellar density maps}\label{dmfromstar}

The second use case is to generate dark matter density fields $\ofield=$\dm conditioned on $\cfield=$\Star. To this end, we only show two cases of \tng with upscaling, having already demonstrated above the generative capability for boxes of the same size as used in training. We therefore use for training the $256^3$ voxel sub-cube of the full \tng run, and further sub-divide into $64^3$ voxel sub-volumes with the same data augmentation strategy as in Section~\ref{gasfromdm}, yielding $\approx 8{,}200$ training samples.
In addition to this baseline setting, we introduce a new test of the generative capabilities across different simulations, in which the model, trained on the small \camels boxes, is applied to the full \tng volume to infer \dm from \Star. This \camels $\rightarrow$ \tng transfer experiment allows us to assess the robustness of the conditional mapping across simulation suites and resolutions. Specifically, we take four of the small \camels simulations, each discretized into $128^3$ voxels, subdivide them into $64^3$ voxel sub-volumes, and apply the same data augmentation procedure to obtain $\approx2{,}600$ individual samples.

In all cases, the data are split into $70\%$ for training and $30\%$ for testing.

\subsubsection{Diffusion model training and validation}
The training procedure is identical for both applications described in sections~\ref{gasfromdm} and~\ref{dmfromstar}. 

At each iteration, the target field, either $\textbf{T}_g$ or $\boldsymbol{\rho}_\text{DM}$, is sampled from the training set and corrupted with Gaussian according to Eq.~\eqref{eq:Gaussian_noising}.  
The optimizer employed for training is AdamW \citep{adamw}, with a batch size of 8. The learning rate schedule adopted is Cosine Annealing Warm Restarts, with the maximum and minimum values for the learning rate set to $3\times10^{-4}$ and $10^{-6}$, respectively, with repeat cycles of 20 epochs. The model is trained for 100{,}000 steps and the model checkpoint that corresponds to the lowest loss is stored.

For the \camels simulations we set $T=400$ for both \dm and \Tg generation, while for the case of \tng, we choose $T=200$ for \Tg generation. For the case of \tng from \camels a timestep of $T=100$ for \dm generation is chosen, as a compromise between generation accuracy (which requires larger $T$) and computational effort (which increaes with larger $T$). 

\subsection{Field Overlap Latent Diffusion method}

\label{subsect:contgen}

As explained above, the diffusion model is trained on small sub-volumes, which for the case of \tng are 1\% of the volume one wishes to generate the target field for. The central problem is how to then combine the generated fields in such a way to obtain an optimal generation for the much larger target volume, that was never seen in during training.  

Simply generating independent sub-volumes --each with the same size of the training data volume-- and then assembling them to generate the target field leads to discontinuities at the cube boundaries, as illustrated in the bottom left panel of Figure~\ref{fig:cont_gen} for the case $\cfield$=\dm and $\ofield=$\Tg. The discontinuities arise because \Tg is sampled from the conditional distribution  $\textbf{T}_g\sim p(T_g|\rho_\text{DM})$ which depends solely on the local voxels \dm sub-volume, without considering the surrounding context. Consequently, the edges of adjacent cubes do not perfectly align. However, the boundary artifacts do not impede the model’s primary objective of capturing the small-scale relationship between dark matter and the temperature field. 

In \LODI, addresses this issue by using overlapping sub-volumes at the generation stage. Each cube being generated shared a boundary region with its neighbors so that latent representations were correlated across overlaps. This ensured that the boundary conditions between sub-volumes in the overlapping region were aligned. 
Although \LODI was shown to work well when the overlapping regions feature localized structures, the approach fails for extended structures that exceed the size of a sub-volume, such as dark matter filaments or \Tg, as demonstrated in the bottom central panel of Figure~\ref{fig:cont_gen}. The second issue of \LODI is the increased runtime: for example, to generate a $256^3$ voxel simulation with \LODI, we would would need to tile a minimum of 512 $32^3$ sub-volumes (this being the training sub-volume size chosen for \LODI).  However, due to the requirement to overlap sub-volumes, the number of forward generation sub-volumes needed increases by 40\% to $729$, with a corresponding increase of compute time. Increasing the size of the training volume would in principle help to mitigate this, but the attention block's computational cost scales quadratically with input dimensions, thus making generation slower rather than faster for larger sizes. Second, because the attention module was not trained on larger spatial volumes, it would need to extrapolate beyond its learned regime, potentially degrading correlation accuracy.
\vspace{\baselineskip}

To address all of the above issue, we introduce here the novel \FOLD method (summarized in Algorithm~\ref{alg:sample}), a generation scheme that builds upon the overlapping–sub-volume strategy while overcoming its key limitations. \FOLD is able to produce large-scale, spatially coherent cosmological fields with higher fidelity and lower computational cost than \LODI, thus enabling efficient synthesis of full-volume realizations without sacrificing small-scale structure detail.

Unlike other approaches that denoise overlapping chunks with some prescription to enforce continuity, \FOLD\ employs a differential sliding window strategy that maintains coherence implicitly throughout the diffusion trajectory, Crucially, this does not use any overlap prescription.
This design reduces generation time by $\approx40\%$ relative to \LODI\ when applied to the \tng simulation (assuming an 8-voxel overlap per side for $64^3$ chunks) even before accounting for the additional overhead of computing overlaps.
Furthermore, \FOLD\ is explicitly designed to respect periodic boundary conditions, which were not enforced in \LODI.

\subsubsection*{\FOLD generation scheme}
To illustrate the \FOLD method, we consider generating a target $\ofield$ field, conditioned on $\cfield$, with a volume of $256^3$ voxels. We first initialize a noise map of the same shape as the target, which corresponds to $\ofield^{t=T}$, where the upper index denotes the step in the denoising process. Next, this cube is partitioned into a $4^3$ grid of non-overlapping chunks, each with $64^3$ voxels -- matching the size of the training samples. We then apply one step of the diffusion denoising model ($t \rightarrow t-1$) to each chunk. Each $64^3$ sub-volume is denoised independently for a single time step according to the learned conditional distribution: 
\begin{equation}
    \tilde{\ofield}^{t-1}_{j} \sim p_{\NNpars}\left( \ofield^{t-1}_j \mid \ofield^{t}_j, \cfield_j^t \right),
\end{equation}
where $j=1, \dots, 4^3$ indexes the sub-volume position and $\cfield_j^t$ is the sub-volume $j$ of the field being conditioned upon, $\cfield$. While the conditioning field is independent of diffusion time, we use the superscript $t$ to denote the voxel shifting corresponding to time $t$ (see below).  The one step denoised subvolumes $\tilde{\ofield}^{t-1}_j$ are first arranged side by side to obtain the intermediate large volume field $\tilde{\ofield}^{t-1}$. We then shift this intermediate field by one voxel along each of the three spatial axes to obtain the updated field $\ofield^{t-1} = \mathrm{Shift}(\tilde{\ofield}^{t-1})$.
The conditioning field must be shifted in an analogous manner to remain consistently aligned with the updated tiling. We therefore denote by $\cfield^{t}$ the conditioning field at the grid orientation used at diffusion step $t$, and obtain $\cfield^{t-1}$ by applying the same one-voxel shift to $\cfield^{t}$, i.e. $\cfield^{t-1} = \mathrm{Shift}(\cfield^{t})$.

The pair $(\ofield^{t-1}, \cfield^{t-1})$ then serves as input to the next denoising step. In this way, at each iteration the entire cube is denoised again using the most recent latent state, but with an offset grid so that different voxel neighborhoods are processed together over successive steps.

\begin{figure*}
    \centering
    \includegraphics[width=0.8\linewidth]{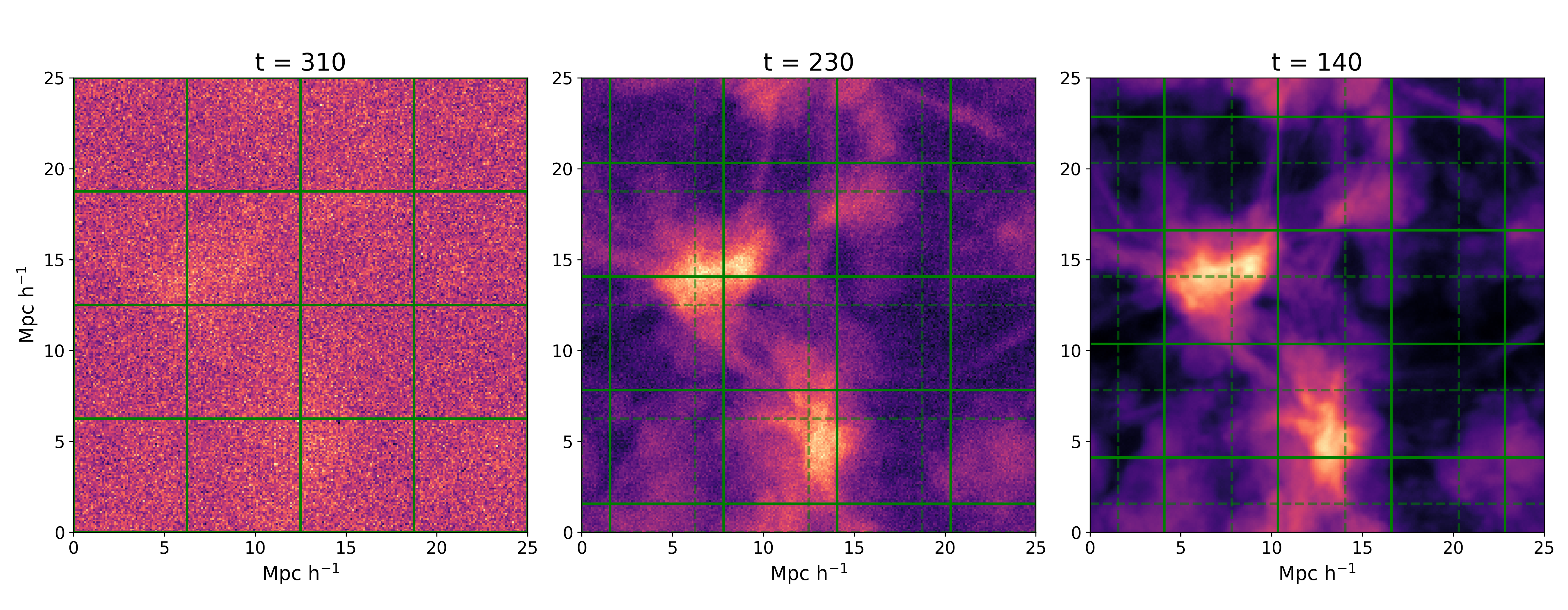}
    \caption{Illustration of the \FOLD overlap latent diffusion method trained on \camels to generated a gas temperature map conditioned on dark matter: example generated gas temperature cube at $t=310$ (left panel), with its sub-volume grid shown in green solid lines. At a later diffusion timestep, $t=230$ (middle panel), the gas field has been denoised, with solid lines showing the current grid and dashed lines the grid location at $t=310$. The process continues until $t=0$ is reached, with the right panel showing another snapshot at $t=140$.}
    \label{fig:sliding_window}
\end{figure*}

This reconstruction and grid shift occurs a total of $T$ times, one for each time step, ensuring that information propagates smoothly across sub-volume boundaries. Since the simulation fields satisfy periodic boundary conditions, the grids also wrap around the edges of the simulation box when shifted. An illustration of our algorithm is shown in~Fig.~\ref{fig:sliding_window}, for the use case in the first line of Table~\ref{tab:fields}:  the image on the left is the full \camels volumes generated at time $t=310$ (about a quarter through the denoising process, as here $T=400$), and the partition of the sub-volumes follows the green grid lines. Some structure begins to emerge from the initial pure noise. After another 80 time steps (central panel), the new grid structure is shown with solid lines, and the old grid is dashed. At $t=140$ step (right image), the grid has further shifted and the noise level has been reduced.

\begin{algorithm}
\caption{\FOLD}
\label{alg:sample}
\begin{algorithmic}[1]
\State $\ofield^T \sim \mathcal{N}(\mathbf{0}, \unitymat)$
    \For{$t = T-1, \dots, 0$}
        \State Partition in sub-volumes
        \For{cube $j \in$ sub-volumes}
        \State $\tilde{\ofield}^{t-1}_{j} \sim p_{\NNpars}\left( \ofield^{t-1}_j \mid \ofield^{t}_j, \cfield_j^t \right)$
    \EndFor
    
    Tile $\tilde{\ofield}^{t-1}_j, \cfield_j^t$ to $\tilde{\ofield}^{t-1}, \cfield^t$  \quad \#{ recreate the original volume size at time $t-1$}
    \State shift tiled $\tilde{\ofield}^{t-1}, \cfield^t$ by 1 voxel along 3 axes with periodic boundary conditions to get $\ofield^{t-1}$ and new shifted $\cfield^{t-1}$
\EndFor
\State \textbf{return} $\ofield^0$
\end{algorithmic}
\end{algorithm}
\vspace{\baselineskip}

The success of \FOLD follows from the fact that the algorithm does not privilege any fixed grid orientation, thereby preventing systematic edge artifacts.
After each shift, every new sub-volume overlaps with several one-voxel-wide planes that were denoised in previous time steps.
These boundary planes are initially slightly out of distribution relative to the interior voxels, but the subsequent denoising pass naturally smooths away such small discontinuities, and adds spatial correlations in those voxels, ensuring global coherence.
Provided that the noise increment per time step is sufficiently small, the diffusion model perceives no abrupt transitions, and the denoiser refines the entire field as a continuous, spatially consistent structure across iterations.

\subsubsection*{Including positional encoding for accurate bispectrum}
\label{bispectrum}

The two-point correlation function or its Fourier counterpart, the power spectrum, is the most widely used statistic to characterize the statistical properties of a cosmological field. However, the power spectrum fully describes the field only if the underlying density fluctuations are a Gaussian random field, which is not the case at later cosmological times~\citep{Bernardeau_2002,sefu06}.To capture deviations from Gaussianity, higher-order statistics are required, the lowest-order of which is the bispectrum -- the Fourier transform of the three-point correlation function. 

For a given density field $\rho$, the bispectrum $B(\mathbf{k_1},\mathbf{k_2},\mathbf{k_3})$ is defined as:
\begin{equation}
    \langle \delta(\mathbf{k_1})\delta(\mathbf{k_2})\delta(\mathbf{k_3}) \rangle = (2\pi)^3 \delta^D(\mathbf{k_1}+\mathbf{k_2}+\mathbf{k_3})B(\mathbf{k_1},\mathbf{k_2},\mathbf{k_3}) \, ,
\end{equation}
where $\delta(\mathbf{k})$ is the Fourier transform of the overdensity field $\delta(\mathbf{x})$ and $\delta^D$ is the Dirac delta function. Since a triangle configuration in $\textbf{k}$ space can also be represented with two wave modes and the angle $\theta$ between them, we will express the  bispectrum a $B( \mathbf{k_1},\mathbf{k_2}, \theta)$.

The bispectrum is an important source of cosmological information: cosmological parameters (like the amplitude of fluctuations, bias of galaxies, neutrino mass, etc.) can be degenerate in terms of the power spectrum, and the bispectrum  helps break these degeneracies, especially in the context of galaxy clustering. In addition, several inflationary scenarios predict the presence of primordial non-Gaussianity \citep{Bartolo_2004}. It is therefore important that the bispectrum of dark matter is predicted accurately by the diffusion process. 

In practice, it is also often useful to consider the \emph{reduced bispectrum}, which isolates the shape dependence of the three point function by factoring
out the overall amplitude set by the power spectrum.
The reduced bispectrum $Q(\mathbf{k}_1,\mathbf{k}_2,\mathbf{k}_3)$ is defined as
\begin{equation}
Q(\mathbf{k}_1,\mathbf{k}_2,\mathbf{k}_3)
=
\frac{
B(\mathbf{k}_1,\mathbf{k}_2,\mathbf{k}_3)
}{
P(k_1)P(k_2)
+
P(k_2)P(k_3)
+
P(k_3)P(k_1)
},
\end{equation}
where $P(k)$ denotes the power spectrum of the overdensity field.
By construction, the reduced bispectrum is less sensitive to the overall normalization of the density field and more directly probes the configuration
dependence of non-Gaussian mode coupling.

\vspace{\baselineskip}

Since the bispectrum and the reduced bispectrum are a function of both the separation between the points and the angle between the wave modes, there exists infinitely many unique triangle configurations. It is therefore impossible to train a model to learn the bispectrum for all possible configurations individually. 
To address this issue, and to enable the model to capture the scale and angular dependence of the bispectrum across all configurations, we supplement the \FOLD input with periodic positional encodings. These provide global phase information that is otherwise inaccessible to purely local convolutional filters of the diffusion model. For each spatial coordinate axis, we construct a positional feature as a sum of sinusoidal functions at multiple frequencies:
\begin{equation} \label{eq:pos_embedding}
   \textbf{p}(\textbf{x})=\displaystyle\sum_{i=0}^n  \alpha_i[\sin(2\pi\textbf{x}f_i) + \beta\cos(2\pi\textbf{x}f_i)] \, ,
\end{equation}
where $\mathbf{x}=(x,y,z)$, and where we work with normalized coordinates, so $x,y,z\in[0,1]$. $\{f_i\}^n_{i=0} = 2^n$ are the frequencies, each dealing with a different scale, thus spanning a hierarchy of scales within the box. $\alpha_i$ are the fixed amplitude weights per frequency bands and $\beta$ ias chosen to add a relative amplitude between the sine and cosine signals, which has shown in our case to improve results. The positional encodings $\textbf{p}(\textbf{x})$ are concatenated to the conditioning field $\cfield$ (in our case, this is \dm), yielding an array of shape $3\times d^3$, where $d$ are the total number of voxels per side. We test the model for different values of the hyperparameter $n$. In principle, the larger the $n$, the more phase information the model has access to. However, it can also confuse the convolutional network since we are adding all frequencies for one coordinate into one channel, making it hard for the network to disentangle the phase information at different scales. For the results of this paper, we use $n=4$, while $\alpha_i=[0.33,0.29,0.25,0.21,0.19]$ and $\beta=1.5$. These values were chosen heuristically based on a limited set of exploratory experiments, in which we varied the relative weighting of low and high frequency components and monitored reconstruction quality. Although we have not systematically explored the full hyperparameter space, these values were found to yield satisfactory results. The coefficients $\alpha_i$ also control the relative strength of the positional encoding signal with respect to the conditioning field $\cfield$. If the signal is too weak (small $\alpha_i$), the model is unable to effectively utilize the positional information. Conversely, excessively large values of $\alpha_i$ inject spurious structure
at the corresponding Fourier modes, which is particularly detrimental given the sparsity of $\cfield$.
The chosen values therefore represent a compromise that allows the network to exploit positional information without overwhelming the physical signal.
We apply this modification of the method on the \tng simulation, where we condition on \Star and generate \dm, and verify the quality of the bispectrum of the generated dark matter map for a few different triangle configurations.

\section{Results}
\label{sec:results}
We evaluate the performance of \FOLD using both unseen test simulations and large-scale generative experiments. We first test the model's performance on an independent unseen realization from the Cosmic Variance subset of the \camels simulation suite, which has simulations with the same cosmology but different initial conditions, with a box size of 25\mpc per side, discretized into $256^3$ voxels. This assesses the model’s generalization to new initial conditions. To demonstrate its capability to scale to larger domains, we then apply \FOLD to the \tng simulation at $z=0$ in two distinct settings:$(i)$ generating the gas temperature field conditioned on the dark matter density, and $(ii)$ generate the dark matter density conditioned on a sparse stellar density field. The \tng volume spans 205\mpc per side and is discretized into $1024^3$ voxels, while the model is trained on only $\approx 1 \%$ of its total volume. For this second experiment, we demonstrate the model performance both when trained on \tng and when trained on \camels, thus showcasing its generalization capabilites beyond the simulation it has been trained on.

\subsection{Gas temperature from dark matter field}

\begin{figure*}
    \centering
    \includegraphics[width=1.0
    \linewidth]{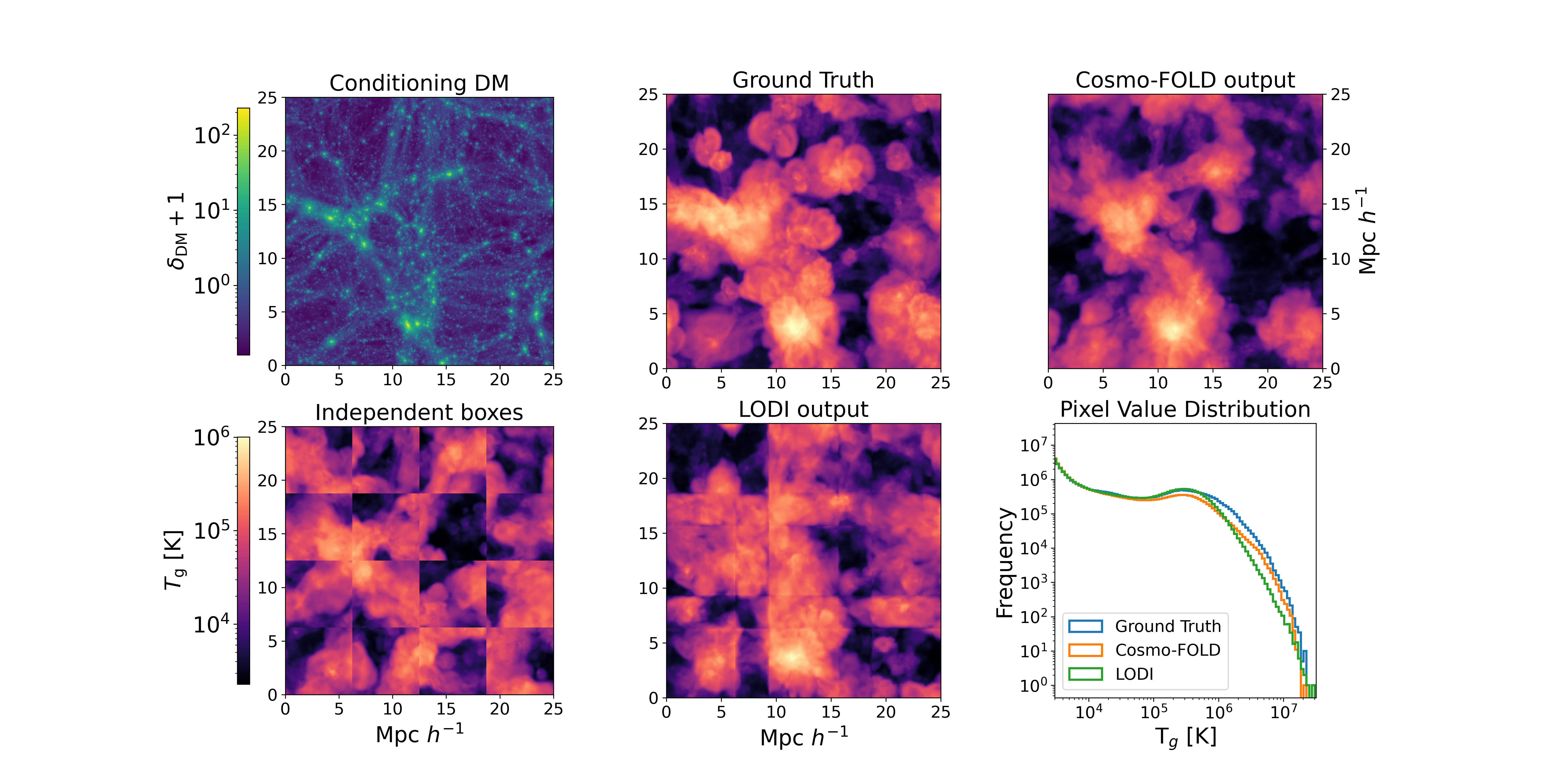}
    \caption{Generated gas temperature map (top right), conditioned on the \dm field (top left) for the \camels simulation, with the target box size equal to the training box size. The true (unseen) temperature map is shown in the top middle panel. For comparison, the bottom left panel shows the output obtained by simply tiling independent sub-boxes, while the bottom middle panel is the output of our previous \LODI method. The histograms of the distribution of pixel values are shown in the bottom right, averaged over 5 realizations of the diffusion models outputs.}
    \label{fig:cont_gen}
\end{figure*}

We first evaluate \FOLD by generating a gas temperature field \Tg conditioned on the dark matter density field \dm, coming from an unseen \camels simulation with the same cosmology and astrophysics as the training simulation, but with different initial conditions. The simulation volume is a cube of side $25$\mpc, that is discretized into $256^3$ voxels. The true (unseen) and \FOLD-generated gas temperature maps are shown in Fig.~\ref{fig:cont_gen} in the top middle and top right panels, respectively. An example generated gas map with simple tiling of independent boxes in shown in the bottom left, exhibiting macroscopic discontinuities at the boundaries of the training boxes, while the output of our previous  \LODI method is shown for comparison in the bottom middle panel. \LODI produces visible edge artifacts at patch boundaries that are absent in the \FOLD method, which delivers seamless, spatially coherent generations and looks faithful to the ground truth.

The bottom right panel shows a comparison of the distribution of voxel values, averaged over 5  generations for the same conditional dark matter field. For temperatures below  $10^6K$, the voxel distribution obtained with \LODI is marginally better than the \FOLD method. However, for larger values \FOLD exhibits a clearly superior performance. This is because in \LODI some regions are generated independently and without overlaps,  which produces more locally accurate results. However, accuracy and overall fidelity is lost due to the subsequent edge effects, which is not the case in our novel approach.

\begin{figure*}
    \centering
    \includegraphics[width=1.0\linewidth]{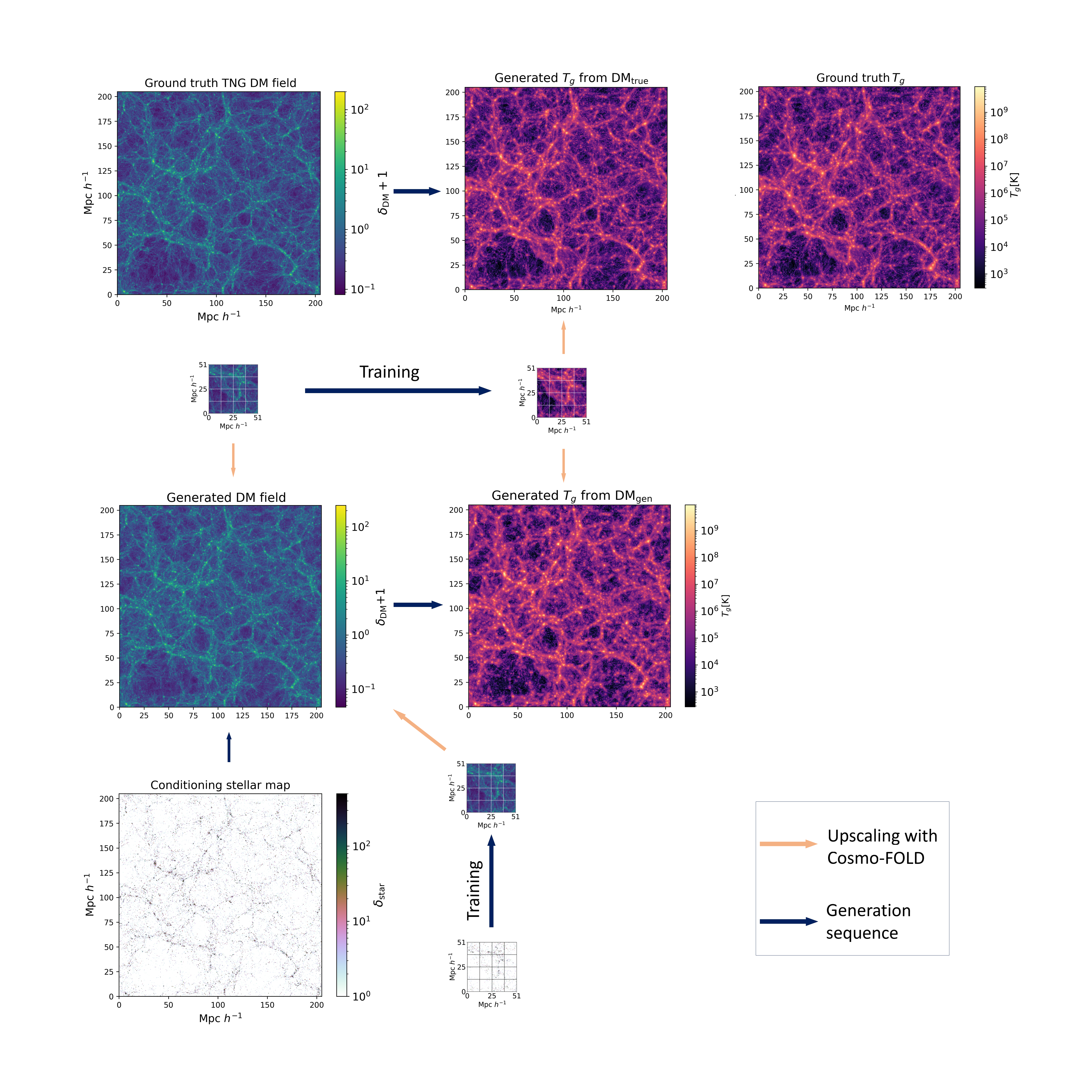}

    \caption{Upscaling to larger volumes via the \FOLD method, applied to generate the target gas temperature field \Tg of \tng (top right), conditioned on the true \dm (top left) or to the \dm field (middle left) generated from the stellar density field (bottom left). Orange arrows depict upscaling, i.e., large volume generation after training on a much smaller volume. The dark blue arrows show the generation direction, pointing from the conditional field to the generated field. }
    \label{fig:upscaling}
\end{figure*}

Next, to evaluate performance on upscaling to larger domains, we apply \FOLD to the \tng simulation (Figure~\ref{fig:upscaling}). Here, the diffusion model is trained and validated on a $256^3$ voxels subset of the data, or $\approx 1.5\%$ of the simulation ($70\%$ of which is for training and rest for validating). The ground truth gas temperature field \Tg is shown in the top right panel, to be compared with two different generation routes obtained with \FOLD. In the first case (top row), the generated \Tg field is conditioned on the {\em true} \dm field of the \tng simulation, while in the second case (middle and bottom row), the conditional \dm field has also been generated using the \FOLD algorithm from the underlying stellar field, as described in section~\ref{sec:dm_from_stars}. 

\begin{figure*}
    \centering
    \includegraphics[width=0.8\linewidth]{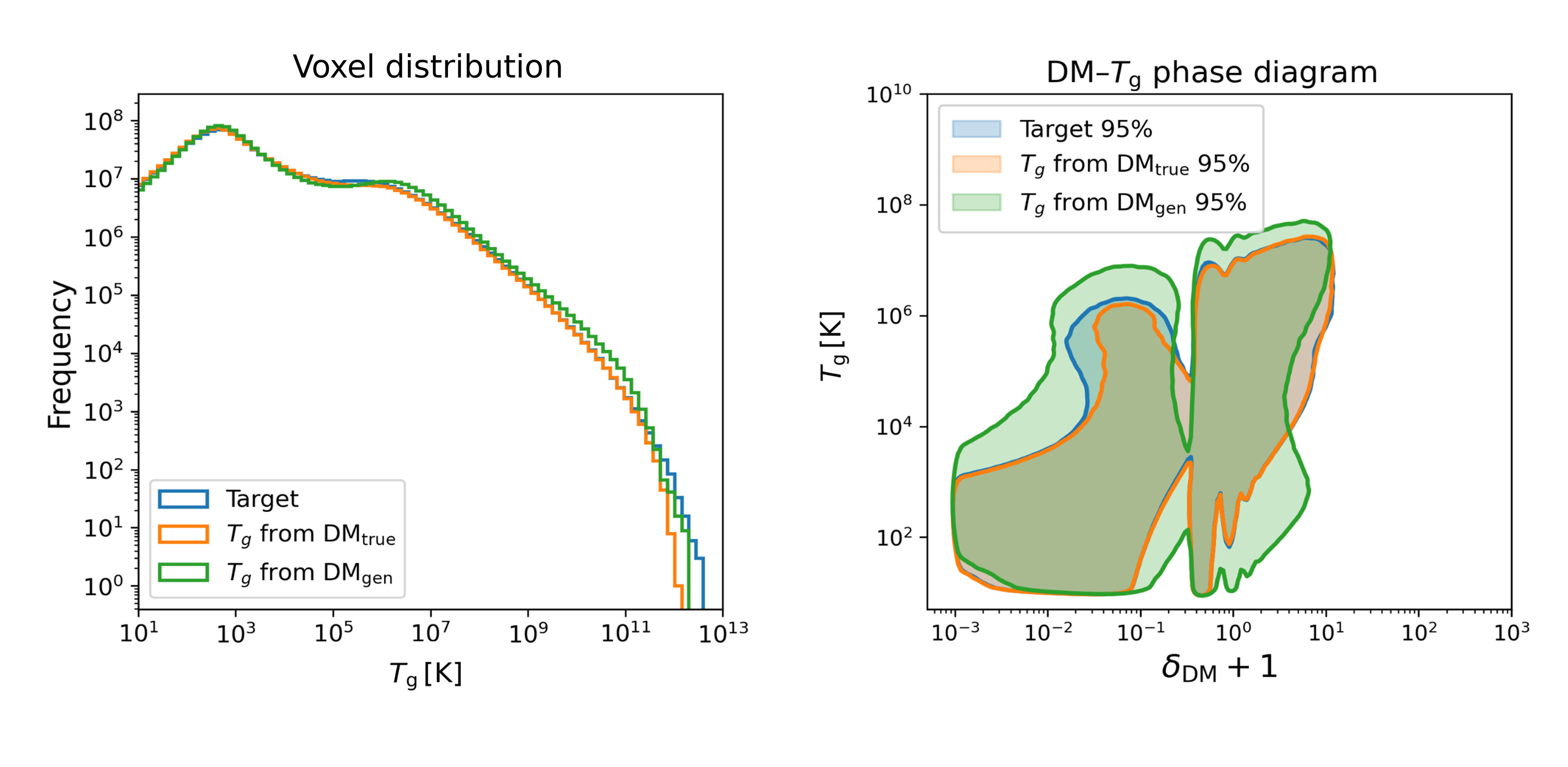}
    \caption{Voxel statistics, comparing the quality of the generated \Tg fields of \tng simulations using the \FOLD pipeline. Left: voxel distribution of the gas temperature field, generated using both the ground–truth \dm field and the \FOLD-generated \dm field. Right: $95\%$ contour levels of the joint \dm-\Tg distribution for the same two generated cases, compared against the ground truth.}
    \label{fig:TNG_DM_mstar_1}
\end{figure*}

We compare the statistical distributions of voxel values using both 1-dimensional histograms and 2-dimensional \dm-\Tg joint distributions in Fig.~\ref{fig:TNG_DM_mstar_1}. The 1D histogram shows that for intermediate \Tg values, ranging from $1\text{K}\leq$\Tg$\leq 10^{11.5}$K, the abundance of voxels is accurately reproduced for both (true and generated) conditioning dark matter maps. At the highest temperatures the model under-predicts rare, extreme-valued voxels, that represent a fraction in the rangr $10^{-6}-10^{-8}$ of the total voxel population, very difficult to capture due to their rarity among the training examples. The 2D contour plot, displaying the $68\%$ and $95\%$ joint probability mass of dark matter density voxel values and \Tg values, shows close agreement between the target and generated contours when the true dark matter map is used for conditioning -- except where large \Tg values correspond to low density dark matter regions, where \FOLD
under-predicts the \Tg values. This is expected because the conditional dark matter density carries limited information about the diffuse feedback heated gas that populate these voxels. The agreement with the ground truth is somewhat reduced when the conditioning map has been itself generated from the stellar map, which is again expected because of the compounding of errors. 

\vspace{\baselineskip}

\subsection{Dark matter from stellar density fields with upscaling} \label{sec:dm_from_stars}

We next apply \FOLD to the inverse problem of reconstructing the dark matter overdensity field $\boldsymbol{\rho}_\text{DM}$, from the corresponding stellar field $\boldsymbol{\rho}_\text{star}$. We first explore the performance of \FOLD at the level of 1- and 2-point statistics, and then add positional encodings to ensure good reconstruction of the bispectrum. Finally, we evaluate the model capabilities to transfer from \camels to \tng. 

\begin{figure*}
    \centering
     \includegraphics[width=0.8\linewidth]{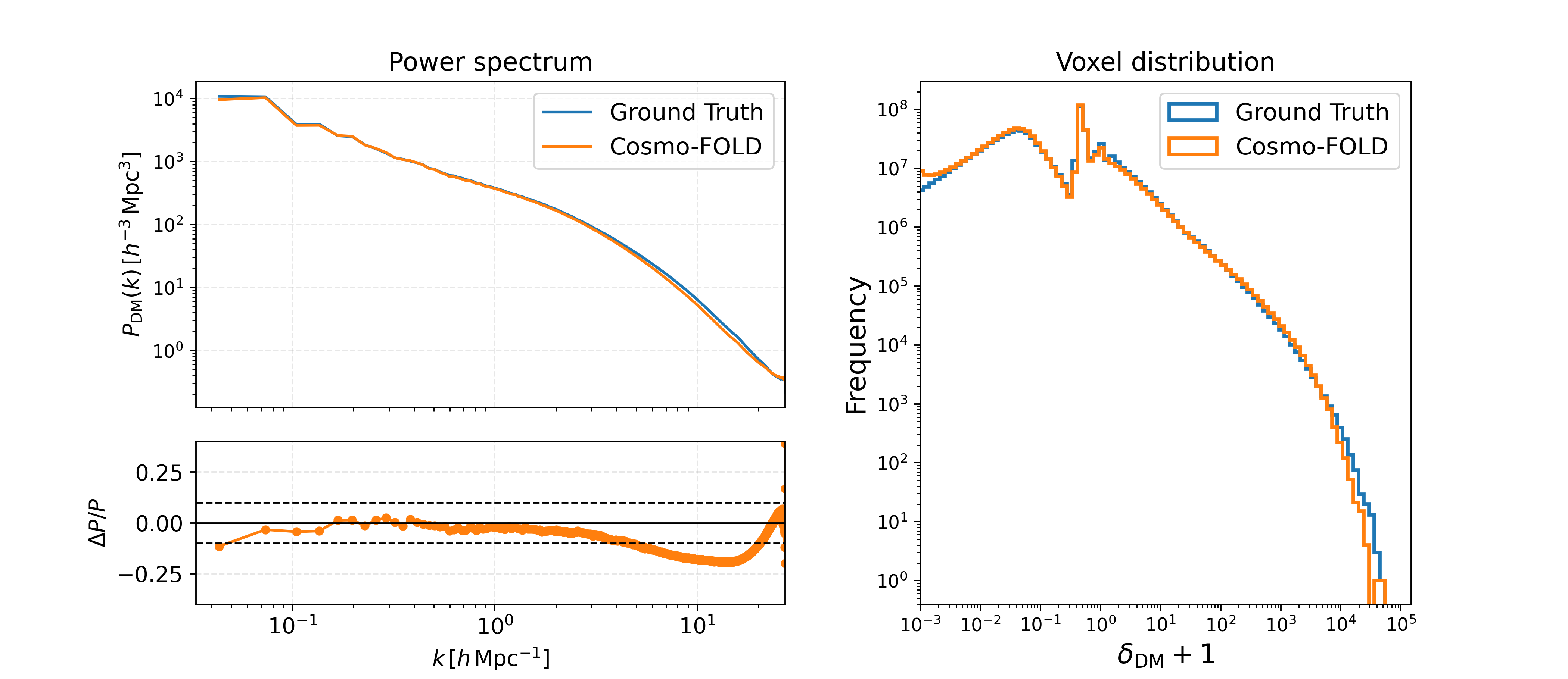}
    \caption{Voxel statistics comparing the quality of the \dm field of the \tng simulation generated from a \tng stellar field with uspcaling. Left: generated dark matter power spectrum together with the relative deviation from the ground truth. Horizontal dashed lines delimit the 10\% level deviation. Right: voxel distribution of the dark matter field generated by \FOLD when conditioned on the stellar density field. }
    \label{fig:TNG_DM_mstar_2}
\end{figure*}

\subsubsection{\tng from \tng}
First, we apply the model to the \tng simulation, generating the \dm field (middle left panel in Fig.~\ref{fig:upscaling}) from the stellar density field (bottom left panel) voxelized into $1024^3$ cells, and covering a cube of side $205$\mpc.
The $256^3$ voxels subvolume used for training and validation $(\approx1.5\%~\text{of the total})$ is shown by the smaller boxes on the bottom right (the grid shows the further subdivison in $64^3$ training volumes). To assess the quality of the generation, we compare the 1 point statistic in the form of a histogram of voxel values and the 2 point correlation statistics in the form of the power spectrum $P_\text{DM}(k)$ of the generated and the ground truth dark matter fields in Fig.~\ref{fig:TNG_DM_mstar_2}, where we also show power spectrum residuals. The power spectrum matches to better than 10\% well into the non-linear regime, increasing only beyond $k\approx 5~h~\text{Mpc}^{-1}$. The reason for the loss of power at the smallest scales is their extreme non-linear nature, which is harder to predict. For the largest scales, the error is a combination of cosmic variance and the approximate correlation for voxel with separations much larger than than the training volume. However, this effect is mild since in our case the wavemode corresponding to the subvolume scale is $k\approx 0.5~h~\text{Mpc}^{-1}$, but the power spectrum does not lose power till $k\approx 0.1~h~\text{Mpc}^{-1}$. We note that the small scale performance can potentially be improved by increasing the total denoising steps $T$, at the cost of increased generation time. 

\subsubsection{Accurate dark matter bispectrum with positional encodings}

Beyond one- and two-point statistics, an important test of generative models for cosmological fields is their ability to reproduce higher order correlations that encode non-Gaussian structure arisign from non-linear gravitational collapse at late times. In particular, the bispectrum of the dark matter density field provides a
sensitive probe of mode coupling induced by non-linear gravitational evolution.
To assess whether our model captures this information, we repeat the generation of the dark matter field from the stellar density field while augmenting the conditioning input with sinusoidal positional encodings
$\mathbf{p}(\mathbf{x})$, as described in Section~\ref{bispectrum}. These encodings provide the network with explicit information about spatial
location within the simulation volume, which can help to break translational degeneracies inherent to patch-based training and improve the modeling of
large scale phase correlations. 

\begin{figure*}
    \centering
    \includegraphics[width=1\linewidth]{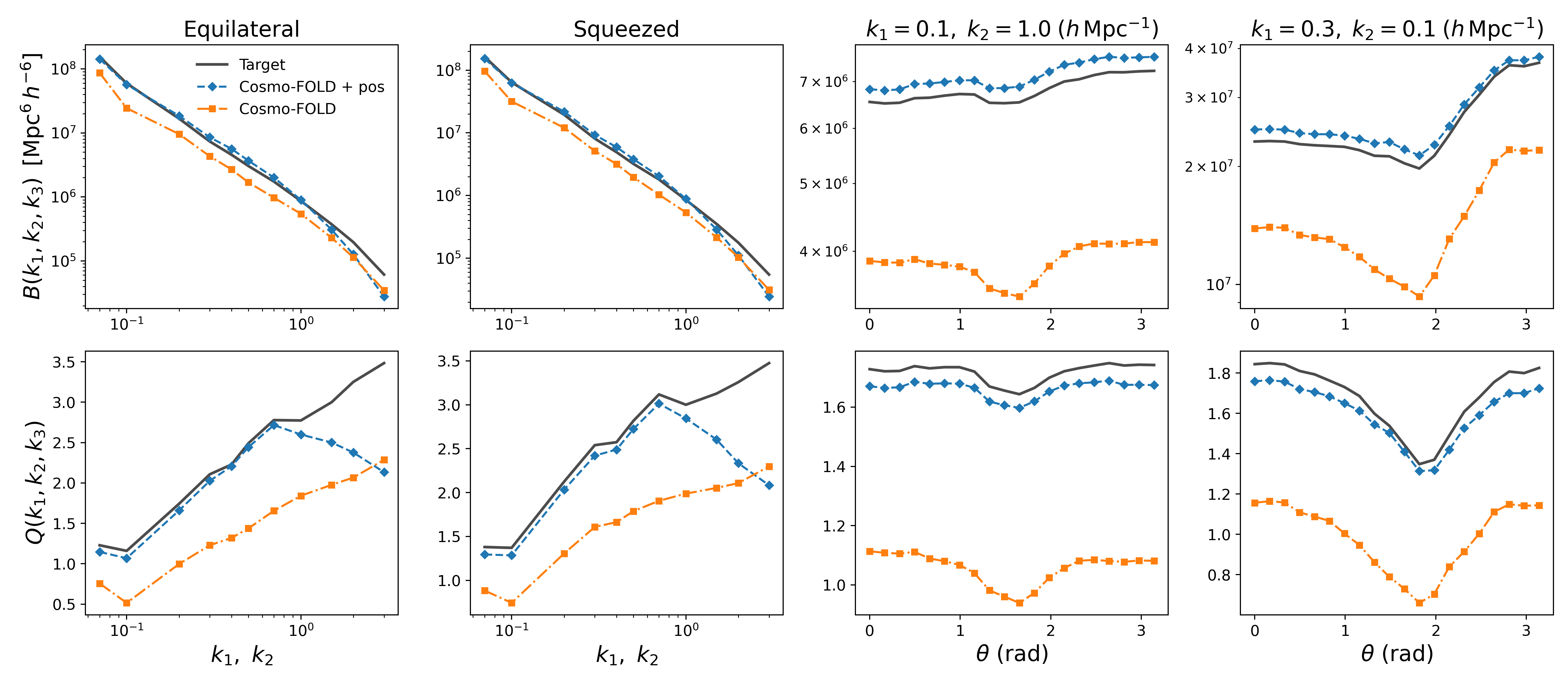}
    \caption{Comparison of four bispectrum configurations (top row) and their corresponding reduced bispectrum (bottom row) between the baseline \FOLD algorithm and the variant augmented with positional embeddings, denoted as \FOLDpos. The equilateral (left column) and squeezed (left-center column) triangle cases are shown as functions of wavenumber, obtained by varying the $k$–values along the corresponding triangle configurations. For the remaining two columns, the wavenumbers are fixed at the values indicated in the panel titles, and the bispectrum is evaluated as a function of the angle $\theta$ between $k_1$ and $k_2$.}
    \label{fig:bispectra}
\end{figure*}

The resulting bispectra and reduced bispectra of the generated dark matter fields are shown in Fig.~\ref{fig:bispectra}, where we compare them to the ground-truth simulation across a range of triangle configurations in Fourier space. In all panels, the blue curves correspond to \FOLD augmented with positional encodings, the orange curves to the baseline \FOLD model without positional encodings, and the black curves to the ground truth.

We present results for four representative triangle configurations (see Section~\ref{bispectrum} for details). The leftmost column corresponds to equilateral configurations with wavevectors $\mathbf{k_1}=\mathbf{k_2}$ and the angle $\theta$ between them as $\pi/3$. The middle-left column shows squeezed configurations with $\mathbf{k_1}=\mathbf{k_2}$, but with $\theta=0$. The middle-right and rightmost columns correspond to fixed-scale configurations with $(\mathbf{k_1},\mathbf{k_2})=(0.1,1.0)h~\text{Mpc}^{-1}$ and $(\mathbf{k_1},\mathbf{k_2})=(0.3,0.1)h~\text{Mpc}^{-1}$.
Across all configurations, we find that the inclusion of positional encodings significantly improves agreement with the true bispectrum. This demonstrates that the model is able to recover non-Gaussian features of the dark matter field that are not constrained by the power spectrum alone. 

\begin{figure*}
    \centering
    \includegraphics[width=0.8\linewidth]{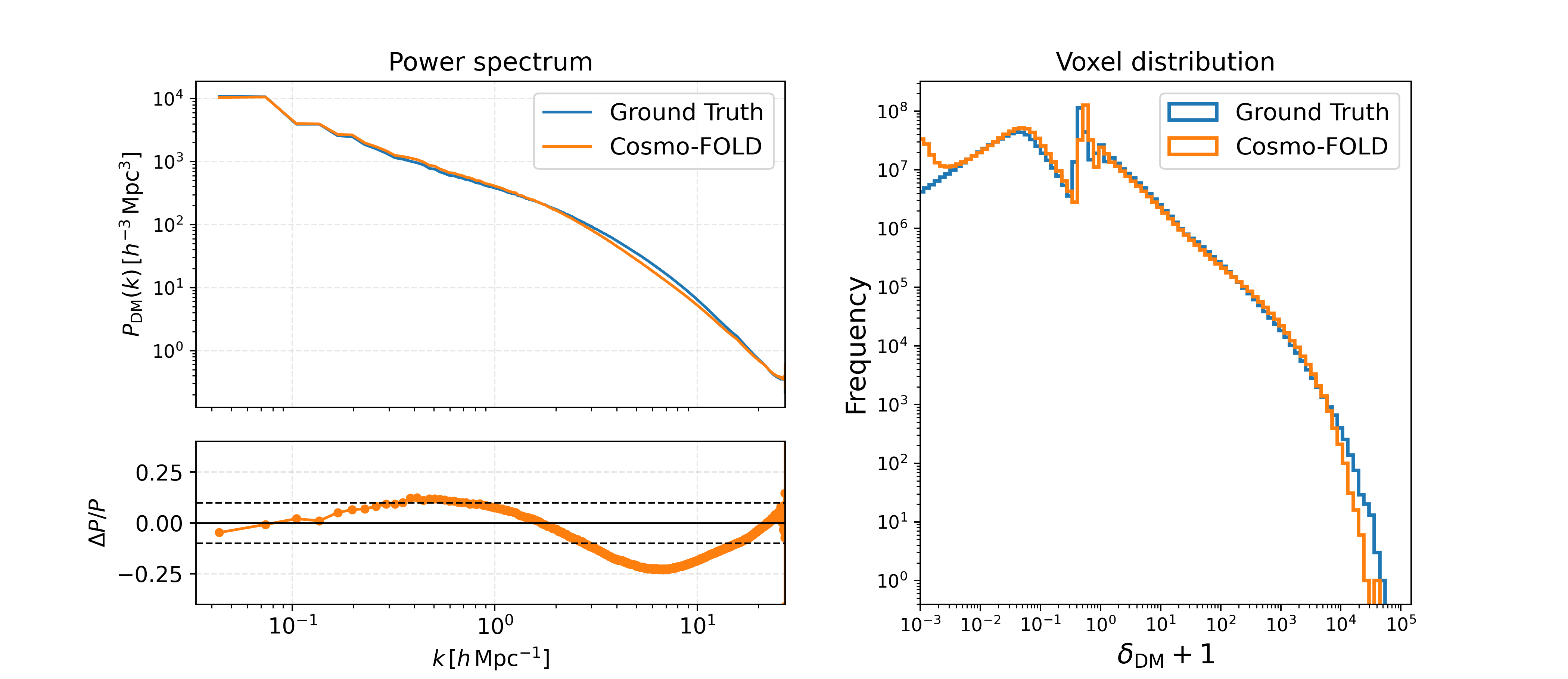}
    \caption{Power spectrum and a histogram plot comparing the 2 point correlation and voxel value distribution between the target and generated \dm field, conditioned on both the \Star and positional encodings and generated using position encodings.}
    \label{fig:bispectra_2}
\end{figure*}

Adding positional encodings does not appreciably reduce the quality of the lower-order statistics. As shown in Fig.~\ref{fig:bispectra_2}, both the one-point distribution and the power spectrum of the generated fields remain in close agreement with the ground truth when positional encodings are added, with only a mild degradation compared to the case without positional encodings (Fig.~\ref{fig:TNG_DM_mstar_2}). We note the presence of additional power at intermediate scales, $k\simeq0.3-1.0~ h~\text{Mpc}^{-1}$, which can be attributed to spurious structure introduced by directly concatenating positional encodings to the sparse stellar field at the voxel level. This additional signal can partially confuse the network and enhance power at these scales. Conversely, power on the largest scales, previously suppressed in the absence of positional encodings, is more accurately recovered. 

Overall, these results indicate that the inclusion of positional information substantially improves the modeling of higher order correlations, while preserving the accuracy of one- and two-point statistics to a high degree.

\subsubsection{\tng from \camels}

\begin{figure*}
   \centering
   \includegraphics[width=0.8\linewidth]{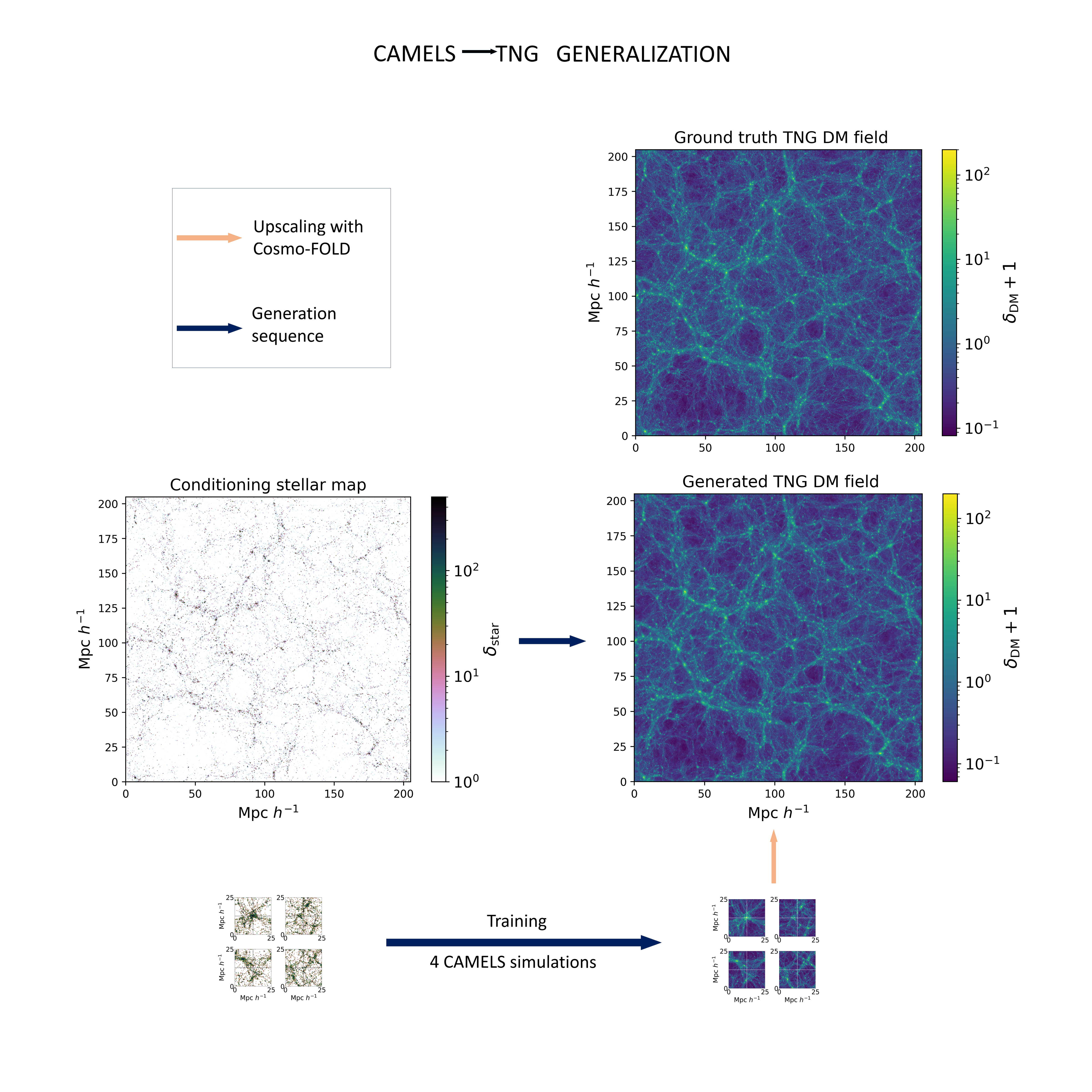}

   \caption{Capability of \FOLD to generate \tng simulations, \dm(bottom right) while being trained on small \camels simulations (bottom four small squares).}
   \label{fig:TNG_CAMELS_illus}
\end{figure*}

Here we revert to the baseline \FOLD model without positional encodings and assess its ability to generate \tng volumes when trained exclusively on small \camels simulations. We train and validate the model on four independent \camels simulations with a comoving box size of side $25$\mpc, generating dark matter overdensity fields conditioned on stellar density fields. Each
simulation is partitioned into $64^3$ voxel subvolumes for training.
Figure~\ref{fig:TNG_CAMELS_illus} illustrates the training and generation pipeline, with the four \camels simulations shown in the bottom row. To quantitatively assess the quality of the generated \tng dark matter field, we compare its power spectrum to that of the ground truth. We note that visually, the generated field looks very similar to the ground truth field, recreating most of the filaments and void, seven if during training, the filaments and structures seen by \FOLD were very different from \tng, and much smaller in size.

\begin{figure*}
   \centering
   \includegraphics[width=0.6\linewidth]{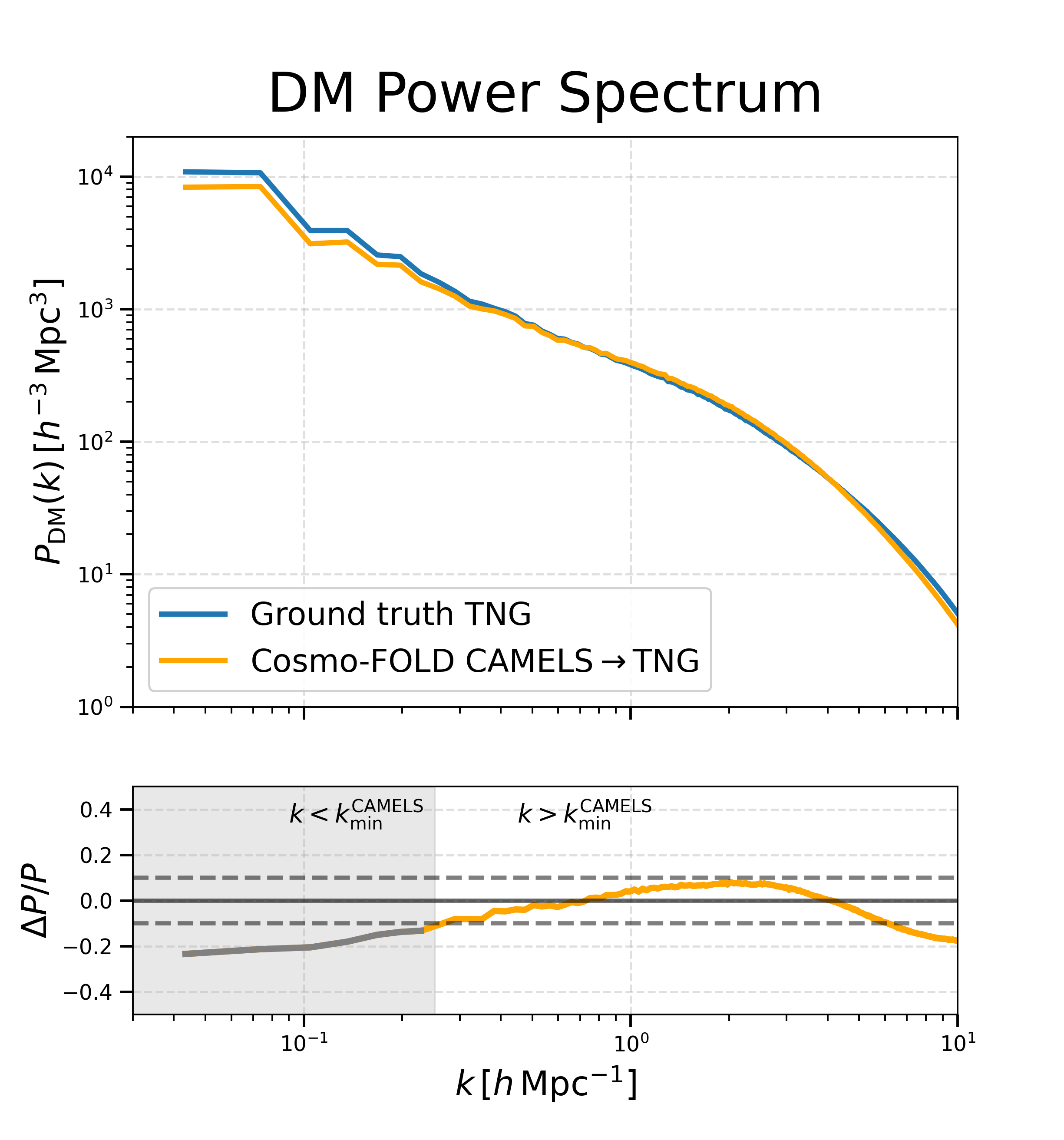}

   \caption{The dark matter power spectrum generated by the \FOLD method, when trained on \camels simulations and upscaled to produce the \tng map, as illustrated in~Fig.\ref{fig:TNG_CAMELS_illus}. The bottom plot shows the relative difference in the power spectrum when compared to the \tng ground truth dark matter spectrum. The gray shaded areas show the scales which contain wavemodes larger than those present in the \camels simulation data. }
   \label{fig:TNG_CAMELS_PS}
\end{figure*}

The resulting power spectrum exhibits two distinct regimes, as shown in Fig~\ref{fig:TNG_CAMELS_PS}. On scales smaller than the \camels box size, the generated field reproduces the true power spectrum to within $10\%$ accuracy, extending into the non-linear regime up to $k \leq 6~h~\mathrm{Mpc}^{-1}$. In contrast, for scales approaching and exceeding the \camels box size, the model exhibits a systematic loss of power, with residuals exceeding the $10\%$ level. This behavior is expected, as \FOLD cannot learn or reconstruct large-scale modes that are absent from the training volumes. As a result, power on scales larger than those present in the training data cannot be faithfully recovered.

\subsection{Computational cost}

\begin{table*}
\centering
\caption{Computational cost associated with the \FOLD use cases presented in this work. All results were obtained on a single NVIDIA A100 GPU with 40\,GB of memory. Reported times correspond to wall-clock generation time. For each generated field, we indicate its physical content, its box side and the number of voxels. We also indicate the method that was used to generate the said volume along with the number of denoising steps used.}
\label{tab:compute_cost}
\begin{tabular}{l l l c c}
\hline
\textbf{Use case} & \textbf{Generated volume} & \textbf{Method} &
\textbf{Denoising steps $T$} & \textbf{Wall-clock time} \\
\hline
\multicolumn{5}{c}{\camels} \\ \hline
Gas temperature from DM & \field{\Tg}{25}{256} & Independent boxes & 100 & $\sim$6 min \\
Gas temperature from DM & \field{\Tg}{25}{256} & \FOLD & 100 & $\sim$6 min \\
Gas temperature from DM & \field{\Tg}{25}{256} & \LODI & 100 & $\sim$11 min \\
\hline
\multicolumn{5}{c}{\tng} \\ \hline
Gas temperature from DM with upscaling & \field{\Tg}{205}{1024} & \FOLD & 200 & $\sim$10 h \\
DM from stellar field with upscaling & \field{\dm}{205}{1024} & \FOLD & 100 & $\sim$5 h \\
Same as above $+$ positional encoding & \field{\dm}{205}{1024} & \FOLDpos & 100 & $\sim$5 h \\
\hline
\multicolumn{5}{c}{\camels $\rightarrow$ \tng transfer} \\ \hline
DM from stellar field with upscaling & \field{\dm}{205}{1024} & \FOLD & 100 & $\sim$5 h \\
\hline

\end{tabular}
\end{table*}

A tabulated summary of the computational time required to produce the various results presented in this paper is provided in Table~\ref{tab:compute_cost}. All results were generated on a single NVIDIA A100 GPU with 40\,GB of memory. We further note that the generation time scales approximately linearly with the number of denoising steps $T$ and can therefore be reduced by decreasing $T$, at the cost of degraded performance on the smallest scales.

\section{Conclusions}
\label{sec:conclusions}
We have presented \FOLD (Field-Overlap Latent Diffusion), a novel framework for generating conditional, arbitrarily large cosmological fields by coherently stitching together outputs from a diffusion model trained on smaller volumes. Building upon the variational diffusion architecture of \LODI, \FOLD enables seamless reconstruction of large scale structure while preserving local physical fidelity.

A summary of all test cases, conditioning setups, and corresponding figures discussed below is provided in Table~\ref{summary_table}.
We demonstrated the performance of \FOLD through two representative and complementary test cases. In the first case, we generated gas temperature fields conditioned on the dark matter density field. This choice highlights both the ability of the diffusion model to synthesize complex baryonic observables and the effectiveness of \FOLD in patching together highly
structured fields without degrading the generative performance of the base model. To further study these results, we compared the generation of a
$256^3$ voxels, $25$\mpc a side \camels simulation using different patching strategies, providing
a direct comparison with simpler inpainting approaches
(Fig.~\ref{fig:cont_gen}).

In the second case, we considered the generation of the \tng dark matter density field conditioned on the stellar density field. We first trained and validated
the model on a small subset of the simulation volume (approximately $1.5\%$ of the full \tng box) and demonstrated that \FOLD is able to reconstruct the full
dark matter field, including filamentary large-scale structures, while recovering the power spectrum to within $10\%$ accuracy up to $k \leq 5~h~\mathrm{Mpc}^{-1}$. By augmenting \FOLD with positional encodings, we further showed that 3-point statistics, such as the the bispectrum and
reduced bispectrum, are reproduced accurately across a range of triangular configurations, with clear improvements over the case without positional information (Fig.~\ref{fig:bispectra_2}). Finally, we demonstrated the transferability of the approach by generating the same \tng dark matter field
using a model trained on small bx of $25$\mpc a side CAMELS simulations, as shown in (Fig.~\ref{fig:TNG_CAMELS_illus}).

We emphasize that the positional encoding strategy explored in this work should be regarded as a proof of principle, demonstrating that the inclusion of explicit spatial information can substantially improve the recovery of higher-order statistics. The mild degradation observed in the power spectrum at intermediate scales arises from directly concatenating positional encodings at the voxel level, which can introduce spurious structure. This effect is not fundamental to the approach and may be mitigated through
more structured methods of incorporating positional information that avoid injecting artificial power.

We also stress that these two test cases are particularly challenging, since the way in which the sub-grid small scale physical processes are coupling small scales to larger environment will be radically different. For example, it is envisaged that galactic feedback is likely to generate large temperature fluctuations and eject hot/warm baryons, along path of least resistance and pollute cosmic voids. On the contrary, the physics of star formation will be sensitive to gas properties well inside virialized haloes, and could be regarded as a more local physical effects. The fact that our method manages to capture both scenarios with reasonable accuracy could be regarded as a great success.

Moreover, we note that the generation times reported in
Table~\ref{tab:compute_cost} can be substantially reduced by decreasing the number of denoising steps $T$ and/or by lowering the target resolution, depending on the requirements of a given application. As an illustrative example, reducing the number of denoising steps to $T=50$ and the output resolution from $1024^3$ to $512^3$ voxels leads to only a modest degradation of small scale features, while reducing the total generation time by approximately a factor of 16. In this configuration, a full field can be generated in approximately
19 minutes, while going from the maximum theoretical resolved wavemode from $k_\text{max}\simeq31~h~\mathrm{Mpc}^{-1}$ to a $k_\text{max}\simeq16~h~\mathrm{Mpc}^{-1}$, which remains higher than the smallest scales that the model can reliably predict.

\vspace{\baselineskip}

Having a generative prescription capable of producing large-scale cosmological fields while preserving small-scale structure provides a powerful route toward creating realistic mock surveys for current and upcoming experiments such as SKA, the Vera C. Rubin Observatory and Euclid.

Beyond the current showcased demonstration, our framework offers several promising applications. A single model, trained once on small scale field realizations, can be used to generate arbitrary large astrophysical fields, such as gas or temperature fields, directly from inexpensive dark matter only simulations. This enables efficient construction of hydrodynamic analogues of large and cheap dark matter N-body runs, without the prohibitive cost of rerunning the more expensive full baryonic simulations each time.

Our method can also reconstruct underlying three-dimensional dark matter density fields from sparse tracer inputs such as galaxy or stellar density maps, providing a flexible tool for comparing alternative dark matter scenarios and connecting theoretical predictions with observations.

Furthermore, the framework can be combined with \texttt{JERALD}~\citep{Rigo_2025}, a Lagrangian deep learning model that enhances approximate N-body simulations to mimic baryonic feedback effects on dark matter fields. Coupling that with our \FOLD pipeline would allow the generation of consistent astrophysical fields across different cosmologies, feedback prescriptions, and initial conditions. Such fast forward modeling capability could be directly integrated into simulation-based inference pipelines, enabling rapid parameter estimation and cross-correlation analyses across multiple tracers, which will be the subject of future work.

\section*{Acknowledgements}
RT acknowledges co-funding from Next Generation EU, in the context of the National Recovery and Resilience Plan, Investment PE1 – Project FAIR ``Future Artificial Intelligence Research''. This resource was co-financed by the Next Generation EU [DM 1555 del 11.10.22]. RT and MV are partially supported by the Fondazione ICSC, Spoke 3 ``Astrophysics and Cosmos Observations'', Piano Nazionale di Ripresa e Resilienza Project ID CN00000013 ``Italian Research Center on High-Performance Computing, Big Data and Quantum Computing'' funded by MUR Missione 4 Componente 2 Investimento 1.4: Potenziamento strutture di ricerca e creazione di ``campioni nazionali di R\&S (M4C2-19)'' - Next Generation EU (NGEU). The Sherwood and Sherwood-Relics simulations used in this work were possible thanks to the Partnership for Advanced Computing in Europe (PRACE), during the 16th Call. The Sherwood-Relics simulations were also performed using time allocated during the Science and Technology Facilities Council (STFC) DiRAC 12th Call. We used the Cambridge Service for Data Driven Discovery (CSD3), part of which is operated by the University of Cambridge Research Computing on behalf of the STFC DiRAC HPC Facility (www.dirac.ac.uk). The DiRAC component of CSD3 was funded by BEIS capital funding via STFC capital grants ST/P002307/1 and ST/R002452/1 and STFC operations grant ST/R00689X/1. Part of the simulations were postprocessed on Ulysses supercomputer at SISSA.
MV and RT are also partially supported by INFN INDARK grant. MV is also supported by the INAF Theory Grant "Cosmological Investigation of the Cosmic Web'.





\bibliographystyle{mnras}
\bibliography{references}







\bsp	
\label{lastpage}
\end{document}